\documentclass[aps,prl,twocolumn,groupedaddress]{revtex4}
\pdfoutput=1
\usepackage{units}
\usepackage{amsmath}
\usepackage{graphicx}
\usepackage{amssymb}

\begin{document}

\title{Non-white frequency noise in spin torque oscillators and its effect
on spectral linewidth}

\author{Mark W. Keller}
\email{mark.keller@boulder.nist.gov}
\author{M. R. Pufall}
\author{W. H. Rippard}
\author{T. J. Silva}
\affiliation{\emph{National Institute of Standards and Technology, Boulder, CO
80305-3328}}

\begin{abstract}
We measure the power spectral density of frequency fluctuations in
nanocontact spin torque oscillators over time scales up to 50~ms.
We use a mixer to convert oscillator signals ranging from 10~GHz
to 40~GHz into a band near 70~MHz before digitizing the time domain
waveform. We analyze the waveform using both zero crossing time stamps
and a sliding Fourier transform, discuss the different limitations
and advantages of these two methods, and combine them to obtain a
frequency noise spectrum spanning more than five decades of Fourier
frequency $f$. For devices having a free layer consisting of either
a single Ni$_{\text{}80}$Fe$_{\text{}20}$ layer or a Co/Ni multilayer
we find a frequency noise spectrum that is white at large $f$ and
varies as \emph{$1/f$} at small $f$. The crossover frequency ranges
from $\approx\unit[10^{4}]{Hz}$ to $\approx\unit[10^{6}]{Hz}$ and
the $1/f$ component is stronger in the multilayer devices. Through
actual and simulated spectrum analyzer measurements, we show that
$1/f$ frequency noise causes both broadening and a change in shape
of the oscillator's spectral line as measurement time increases. Our
results indicate that the long term stability of spin torque oscillators
cannot be accurately predicted from models based on thermal (white)
noise sources. 
\end{abstract}

\thanks{Contribution of NIST; not subject to U.S. copyright.}


\maketitle

\section{Introduction}

In a spin torque oscillator (STO), a direct current passing through
a reference magnetic layer becomes spin-polarized and transfers angular
momentum to a second magnetic layer that is excited into steady-state
oscillation. The oscillating magnetization causes an oscillating device
resistance, through either the giant magnetoresistance effect or the
tunneling magnetoresistance effect, which in combination with the
bias current generates an oscillating voltage as the output signal.
Interest in potential applications of STOs in integrated microwave
circuits is driven by their rapid frequency tunability, small size
($\lesssim\unit[100]{nm})$, and compatibility with standard semiconductor
processing techniques. Recent reviews cover both the physics \cite{Ralph:2008pr}
and possible applications \cite{Silva:2008xd,Katine:2008sh} of STOs
and other devices based on spin torque effects.

For any oscillator, noise is both an important figure of merit for
applications and a useful probe of internal physical processes. Previous
models of STO noise \cite{Tiberkevich:2007fc,Kim:2008kx,Kim:2008zr,Tiberkevich:2008ys,Kudo:2009fk,SilvaKeller_preprint2010}
have considered noise driven by a thermal source having a white power
spectral density (PSD). Perhaps because most experiments on STOs have
used a spectrum analyzer (SA) to measure signals in the frequency
domain, most previous theoretical work has focused on how frequency
noise determines the width of the spectral line. For purely white
frequency noise, the relation is straightforward: a constant PSD $S_{\textrm{wh}}$
gives a Lorentzian spectral line whose full width at half maximum
is simply $\Delta\nu_{\textrm{wh}}=\pi S_{\textrm{wh}}$ \cite{Lesage:1979dq,Lax:1967cr}.
The situation is more complicated when the frequency noise is not
white. Colored noise in STOs can occur both at high frequencies, due
to an intrinsic relaxation rate that suppresses rapid frequency fluctuations
\cite{Tiberkevich:2008ys,SilvaKeller_preprint2010}, and at low frequencies
as we demonstrate here. The stability of an oscillator cannot be described
by a single number such as linewidth when its frequency noise is colored,
and a measurement of the noise spectrum is required to accurately
predict oscillator performance in specific applications and to test
models of the physical origin of the noise.

In the next section we present generic equations for an oscillator
in the time domain and introduce the various PSDs we use here. Then
we describe our STO devices and our measurement techniques, including
the use of a mixer to facilitate the measurement of signals well above
10 GHz. Next we describe two methods (both employing standard digital
signal processing techniques) for computing the PSD of frequency fluctuations
from the voltage waveform of the STO. While each method has different
bandwidth limitations, when combined they yield noise spectra extending
over more than five decades of Fourier frequency $f$. We present
these combined spectra for two types of STOs, both of which show $1/f$
frequency noise below $f\approx\unit[1]{MHz}$. Finally, we connect
the frequency noise with SA measurements. The measured linewidth is
larger than the value implied by the white part of the frequency noise
spectrum and it increases with measurement time, effects that have
been seen in semiconductor lasers having $1/f$ frequency noise. The
line shape also changes, becoming more Gaussian at long measurement
times, and we discuss how this affects the interpretation of STO line
shape measurements.

\section{Time Domain Oscillator Model}

The voltage output of a generic oscillator can be written as \begin{equation}
V(t)=\left[{V_{0}+\epsilon(t)}\right]\sin\left[{2\pi\nu_{0}t+\phi(t)}\right],\label{eq:GenericOscil}\end{equation}
where $\epsilon(t)$ is the deviation from the nominal amplitude $V_{0}$,
$\nu_{0}$ is the nominal frequency, and $\phi(t)$ is the deviation
from the nominal phase $2\pi\nu_{0}t$. From the total phase \begin{equation}
\theta(t)\equiv2\pi\nu_{0}t+\phi(t)\label{eq:TotalPhase}\end{equation}
we define an instantaneous frequency \begin{equation}
\nu(t)\equiv\frac{1}{2\pi}\frac{d\theta}{dt}=\nu_{0}+\frac{1}{2\pi}\frac{d\phi}{dt}.\label{eq:InstanFreq}\end{equation}
From this equation, it is clear that phase and frequency are equivalent,
not independent, ways of representing oscillator fluctuations.

Oscillator noise is commonly expressed in terms of the PSD %
\footnote{As is common for real signals \cite{Press:2007ek}, we define the
PSD over positive frequencies only and normalize so that its integral
over these frequencies gives the variance of the signal.%
} of $V(t)$, $\phi(t)$, or $\nu(t)$, which we denote as $S_{V}(\nu)\;\unit{[V^{2}/Hz]}$,
$S_{\mathrm{\phi}}(f)\;\unit{[rad^{2}/Hz]}$, and $S_{\nu}(f)\;\unit{[Hz^{2}/Hz]}$.
The units of each PSD are given in square brackets and $f$ is Fourier
frequency. Note that $S_{V}(\nu)$, the quantity measured by an SA
with swept frequency $\nu$, includes amplitude noise that does not
appear in the other two PSDs. These PSDs can be measured in various
ways; we will describe the methods we use below. We will also make
use of the Fourier identity\begin{equation}
S_{\nu}(f)=f^{2}S_{\phi}(f),\label{eq:Snu_Sphi_identity}\end{equation}
which follows from the fact that frequency is the time derivative
of phase.

\section{Experimental Methods}

\begin{figure}
\begin{centering}
\includegraphics[width=8.5cm]{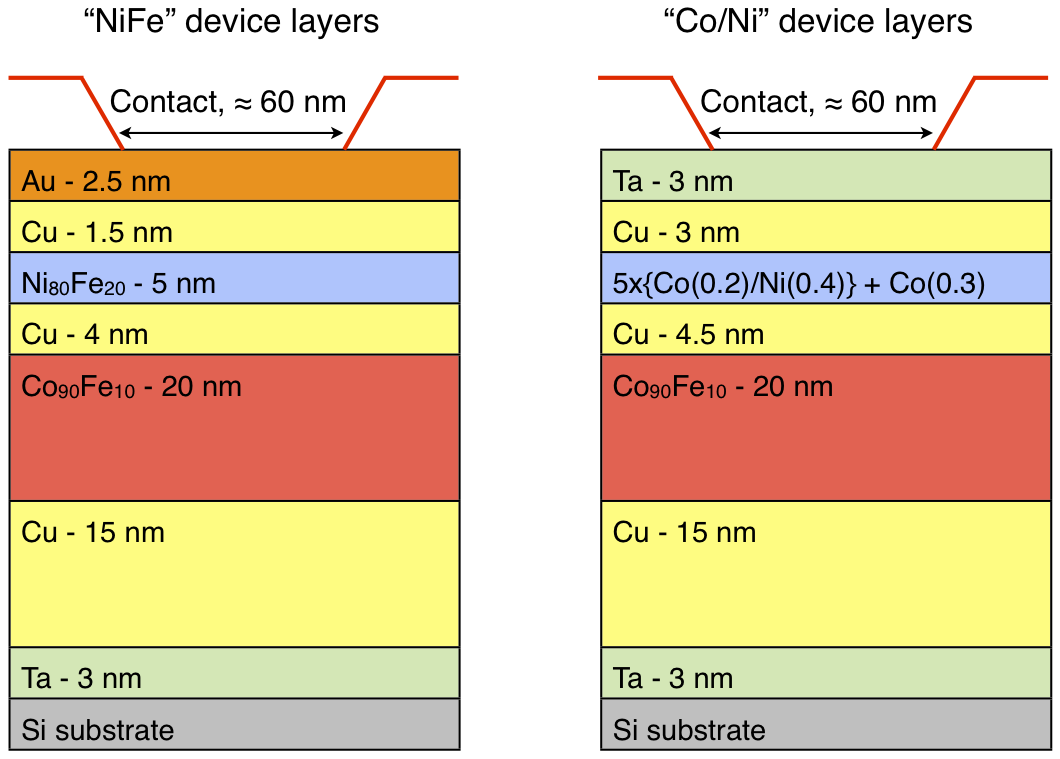}
\par\end{centering}

\caption{(Color online) Spin valve layers used for the NiFe and Co/Ni STO nanocontact
devices. In each case, the layers extend laterally over a mesa of
several $\mu\textrm{m}$ on a side. The current that generates the
spin torque effect flows through a metallic nanocontact to the top
layer and returns through the bottom Cu layer.}

\label{Fig:Spin_valve_layers}
\end{figure}

Our nanocontact STOs consist of a laterally extended spin valve structure
and a metallic contact of nominal diameter 60~nm to 70~nm. Figure~\ref{Fig:Spin_valve_layers}
shows the two types of structures we used. Both structures have a
thick layer of Co$_{\text{}90}$Fe$_{\text{}10}$ that serves as the
reference layer. We use {}``NiFe'' to label devices whose free layer
consists of $\unit[5]{nm}$ of Ni$_{\text{}80}$Fe$_{\text{}20}$
and {}``Co/Ni'' to label devices whose free layer consists of a
multilayer of Co and Ni. With no applied magnetic field, the free
layer magnetization of the NiFe devices lies in the film plane, whereas
that of the Co/Ni devices lies perpendicular to the film plane due
to interfacial anisotropies intrinsic to the multilayer \cite{Daalderop:1992dp}.
The NiFe devices are from the same wafer as those used in \cite{Pufall:2007hb}
and the Co/Ni devices are from the same wafer as those used in \cite{Rippard:2010uq}.

We used microwave probes to contact the devices, which were at room
temperature. The high-frequency STO output, after separation from
the bias current by a bias tee, passed through an amplifier with a
power gain of 36 dB before entering an SA. We measured more than a
dozen devices from four different wafers, at a variety of applied
magnetic fields and bias currents, and we observed non-white frequency
noise in all cases. Here we present representative data from one NiFe
device and one Co/Ni device. The NiFe device was measured with a magnetic
field $\mu_{0}H_{0}=\unit[1.0]{T}$ applied at an angle of $80^{\circ}$
from the film plane and a bias current $I_{\textrm{b}}=\unit[12.1]{mA}$,
for which $\nu_{0}=\unit[13.2]{GHz}$. The Co/Ni device was measured
with $\mu_{0}H_{0}=\unit[1.2]{T}$ applied at $85^{\circ}$ from the
film plane and $I_{\textrm{b}}=\unit[7.5]{mA}$, for which $\nu_{0}=\unit[36.5]{GHz}$.
We chose these conditions to illustrate how devices with nearly the
same SA linewidth ($\approx\unit[11]{MHz}$ in this case) can have
significantly different frequency noise spectra.

\begin{figure}[th]
\begin{centering}
\includegraphics[width=8.5cm]{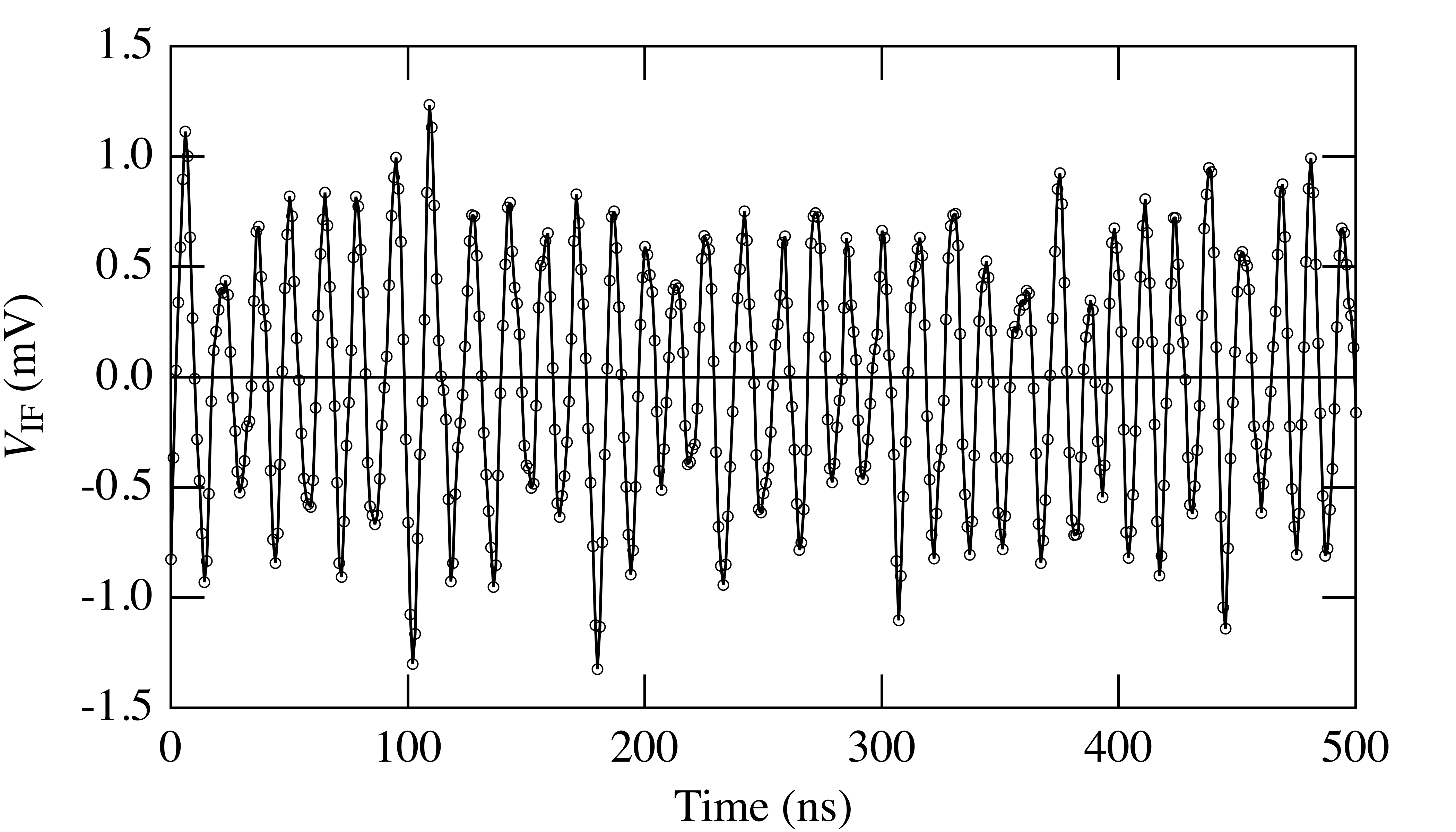}
\par\end{centering}

\caption{A short section of the IF waveform $V_{\textrm{IF}}(t)$ for the NiFe
device. The amplitude variations seen in these data are dominated
by amplifier noise.}

\label{Fig:Vif_vs_Time_NiFe}
\end{figure}

Our technique for time domain measurements was motivated by a desire
to measure STO signals well above 10~GHz using readily available
commercial instruments. We used the intermediate frequency (IF) output
of the mixer in an SA, with a fixed local oscillator (LO) frequency,
to translate the input signal $V(t)$ to an IF signal $V_{\textrm{IF}}(t)$
centered at $\unit[70]{MHz}$. Ignoring the negligible phase noise
of the LO, a perfect sine wave $V(t)$ would appear at the IF output
as a perfect $\unit[70]{MHz}$ sine wave, whereas the frequency or
phase fluctuations in an actual $V(t)$ appear as corresponding fluctuations
in $V_{\textrm{IF}}(t)$. The advantage of the IF measurement technique
is that STO signals for any $\nu_{0}$ within the range of the SA
are translated to a common, lower frequency at which digitization
is straightforward. A disadvantage is that the limited IF bandwidth
prevents the measurement of signals with large linewidths. For the
data presented here, the available IF band was approximately $\unit[(70\pm20)]{MHz}$
and we were limited to STO signals with linewidths $\lesssim\unit[30]{MHz}$.
We used an oscilloscope to digitize $V_{\textrm{IF}}(t)$ at $10^{9}$
samples per second and to apply a digital $\unit[150]{MHz}$ lowpass
filter to reduce preamplifier and oscilloscope noise. Figure~\ref{Fig:Vif_vs_Time_NiFe}
shows a portion of the resulting waveform. The available oscilloscope
memory limited the duration of each filtered waveform to $\leq\unit[50]{ms}$.

\section{Data Analysis Methods}

In this section we describe two different methods for obtaining $S_{\nu}(f)$
from $V_{\textrm{IF}}(t)$. We discuss limitations and averaging considerations
in some detail and we show that the two methods agree over their common
range of $f$. Beyond the points we highlight here, many textbooks
and other sources (e.g., \cite{Press:2007ek}) contain details of
the digital signal processing techniques involved. In the next section
we present composite spectra obtained by combining the two methods
in order to cover a broader range of $f$ than is possible with either
method alone while preserving the benefits of averaging.

\begin{figure}[th]
\begin{centering}
\includegraphics[width=8.5cm]{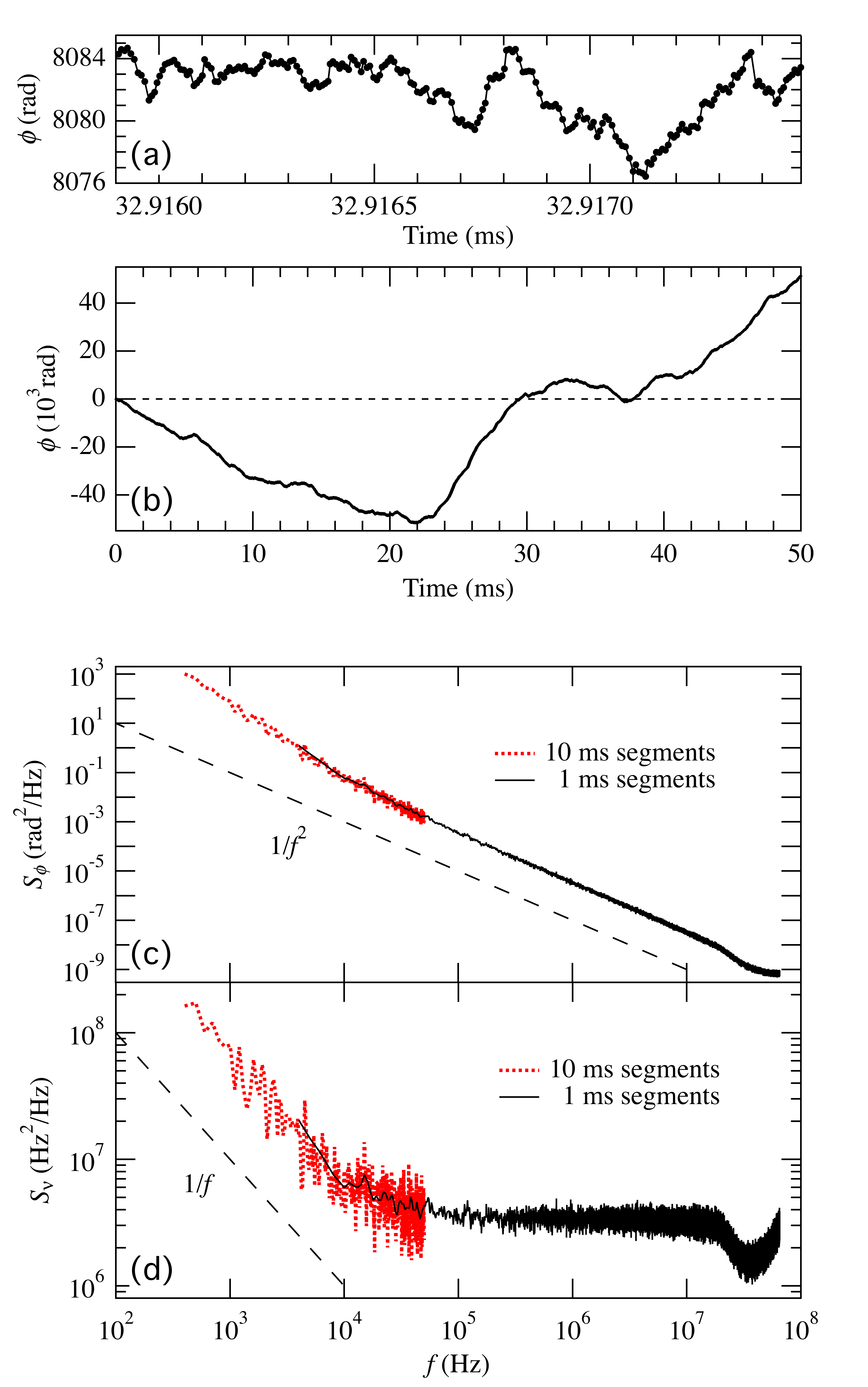}
\par\end{centering}

\caption{(Color online) Zero crossing method for computing $S_{\mathrm{\nu}}(f)$
from $V_{\textrm{IF}}(t)$. (a) Short section of $\phi(t)$ from zero
crossings of $V_{\textrm{IF}}(t)$ for the NiFe device. (b) $\phi(t)$
for the entire $\unit[50]{ms}$ IF waveform. (c) Average $S_{\mathrm{\phi}}(f)$
computed from segments of $\phi(t)$ to reduce scatter in the PSD.
(d) $S_{\mathrm{\nu}}(f)$ computed using Eq.~\eqref{eq:Snu_Sphi_identity}.
For both (c) and (d), dotted curves are the average PSD from 9 half-overlapping,
$\unit[10]{ms}$ segments (points for $f>\unit[5\times10^{4}]{Hz}$
are not shown) and solid curves are the average PSD from 99 half-overlapping,
$\unit[1]{ms}$ segments. $S_{\mathrm{\phi}}(f)$ flattens near $\unit[10^{-9}]{rad{}^{2}/Hz}$
due to the noise floor for the zero crossing method set by amplifier
noise in our setup \cite{Keller:2009ey}, which causes the upturn
in $S_{\mathrm{\nu}}(f)$ at large $f$. Dashed lines are visual guides
indicating $1/f^{2}$ and $1/f$ power law spectra.}

\label{Fig:SnuZeros_NiFe}
\end{figure}

The first analysis method is the ``zero crossing'' method. As described
in \cite{Keller:2009ey}, we obtain a value of the oscillator phase
each time $V_{\textrm{IF}}(t)$ crosses zero, which yields $\phi(t)$
as shown in Figs.~\ref{Fig:SnuZeros_NiFe}(a) and \ref{Fig:SnuZeros_NiFe}(b)
for the NiFe device. Because the values of $\phi$ are not equally
spaced in time, one should in principle estimate the PSD using an
algorithm such as the Lomb periodogram \cite{Press:2007ek} that is
suitable for a time series with irregular spacing. However, in practice
the variations in spacing are sufficiently small that we find no significant
difference between the Lomb PSD and that obtained by assuming uniform
spacing and applying conventional PSD algorithms. Thus for the analysis
presented here we have replaced the actual time stamps in each $\phi(t)$
trace with values separated by the mean spacing for that trace. To
reduce scatter in the spectra \cite{Press:2007ek}, we averaged PSDs
computed from half-overlapping segments of $\phi(t)$, each multiplied
by a Hann window, to obtain $S_{\mathrm{\phi}}(f)$ shown in Fig.~\ref{Fig:SnuZeros_NiFe}(c).
We chose segment lengths of $\unit[1]{ms}$ and $\unit[10]{ms}$ to
balance the tradeoff between averaging and frequency resolution. We
found it necessary to omit the lowest three frequency bins from $S_{\mathrm{\phi}}(f)$
to obtain results that are independent of segment length; thus for
$\unit[10]{ms}$ segments $S_{\mathrm{\phi}}(f)$ begins at $f=\unit[400]{Hz}$
rather than $(\unit[10]{ms})^{-1}=\unit[100]{Hz}$. (Spurious effects
arise when there is a large difference between the initial and final
values of $\phi$ for a segment. Consider a segment where $\phi(t)$
varies linearly: multiplying a line by a window that falls to zero
at each end will create artificially large Fourier components near
the inverse segment length.) Finally, we use Eq.~\eqref{eq:Snu_Sphi_identity}
to obtain $S_{\mathrm{\nu}}(f)$ for the zero crossing method, with
the result shown in Figure~\ref{Fig:SnuZeros_NiFe}(d).

The rolloff in $S_{\mathrm{\nu}}(f)$ beginning at $f\approx\unit[20]{MHz}$
qualitatively resembles the expected rolloff due to the intrinsic
relaxation rate at which an STO returns to its stable precessional
orbit after a fluctuation \cite{SilvaKeller_preprint2010}. However,
in our case this rolloff is due to the limited bandwidth of the IF
output (it occurs near $\unit[20]{MHz}$ in all our data) and does
not reflect an intrinsic time scale of the STO. For the rest of this
paper we show $S_{\mathrm{\nu}}(f)$ from the zero crossing method
only for $f\leq\unit[20]{MHz}$.

\begin{figure}
\begin{centering}
\includegraphics[width=8.5cm]{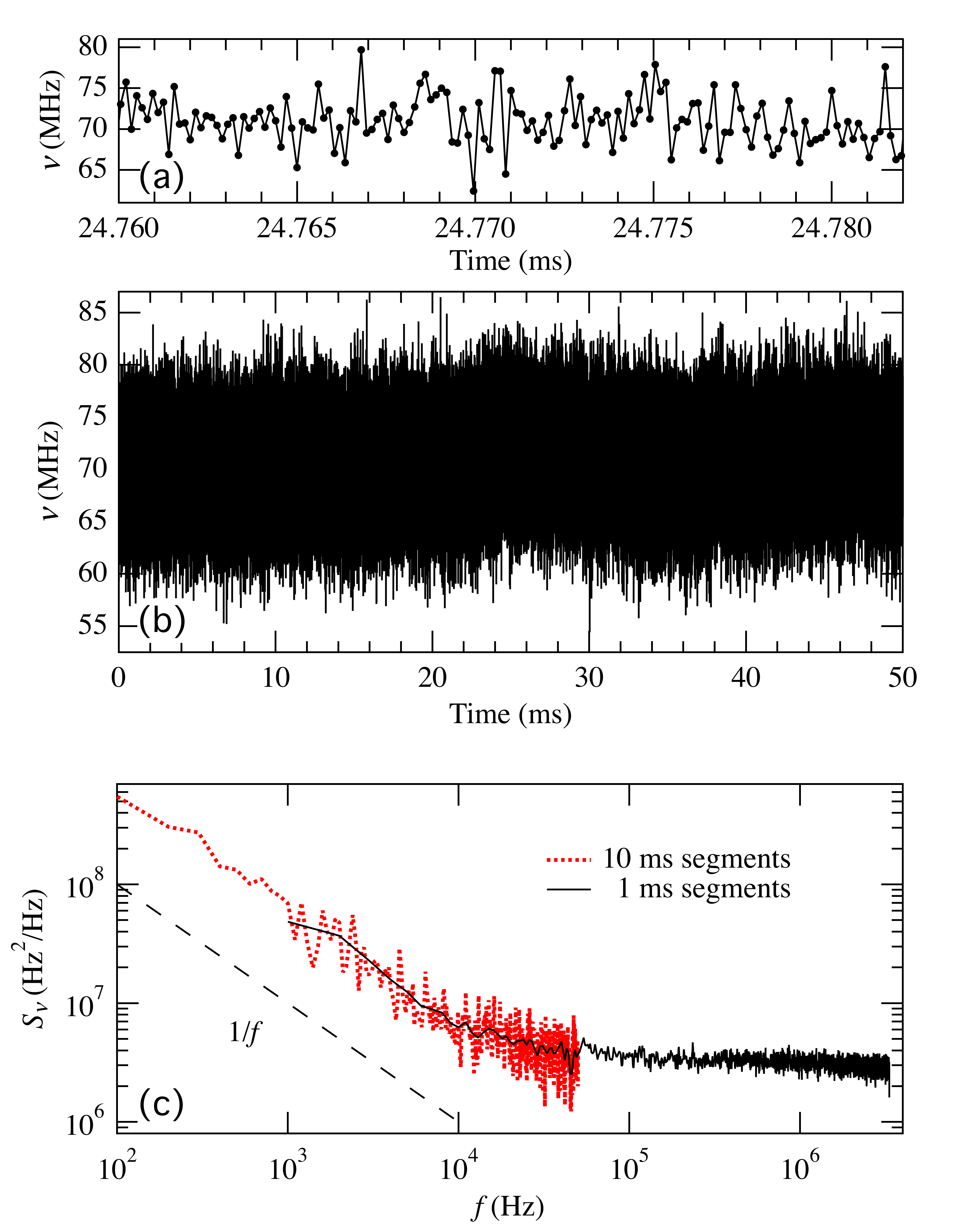}
\par\end{centering}

\caption{(Color online) Sliding DFT method for computing $S_{\mathrm{\nu}}(f)$
from $V_{\textrm{IF}}(t)$. (a) Short section of $\nu(t)$ from DFTs
of half-overlapping segments of $V_{\textrm{IF}}(t)$ for the NiFe
device. (b) $\nu(t)$ for the entire $\unit[50]{ms}$ IF waveform.
(c) Average $S_{\mathrm{\nu}}(f)$ computed from segments of $\nu(t)$
to reduce scatter in the PSD. Dotted curve is for 9 half-overlapping,
$\unit[10]{ms}$ segments (points for $f>\unit[5\times10^{4}]{Hz}$
are not shown). Solid curve is for 99 half-overlapping, $\unit[1]{ms}$
segments. Dashed line is a visual guide indicating a $1/f$ power
law spectrum. }

\label{Fig:SnuDFT_NiFe}
\end{figure}

The second analysis method is the ``sliding DFT'' method. We compute
the discrete Fourier transform (DFT) of half-overlapping segments
of $V_{\textrm{IF}}(t)$ and fit a Lorentzian peak to the DFT to determine
the center frequency for each segment %
\footnote{This analysis should not be confused with that done by a DFT spectrum
analyzer. We take only the center frequency from each segment, not
a complete power spectrum of $V_{\textrm{IF}}(t)$, which is why our
result is independent of amplitude fluctuations ($\epsilon(t)$ in
Eq.~\eqref{eq:GenericOscil}). Amplitude modulation of $V_{\textrm{IF}}(t)$
would create sidebands but would not affect the center frequency of
the main peak.%
}. This yields a trace of $\nu(t)$, as shown in Figs.~\ref{Fig:SnuDFT_NiFe}(a)
and \ref{Fig:SnuDFT_NiFe}(b) for the NiFe device. The shortest segment
that gave reliable results was $\unit[300]{ns}$, which for half-overlapping
segments gives a value of $\nu$ every $\unit[150]{ns}$. We then
compute $S_{\mathrm{\nu}}(f)$ for the DFT method from $\nu(t)$,
again averaging the PSDs of half-overlapping, Hann-windowed segments.
Since the relative excursions in $\nu(t)$ are much smaller than those
in $\phi(t)$, there are no spurious effects for long segments and
$S_{\mathrm{\nu}}(f)$ is independent of segment length over the entire
range of $f$. We could easily obtain $S_{\mathrm{\phi}}(f)$ for
the DFT method using Eq.~\eqref{eq:Snu_Sphi_identity}, but the features
of interest here are more easily seen in $S_{\mathrm{\nu}}(f)$.

\begin{figure}
\begin{centering}
\includegraphics[width=8.5cm]{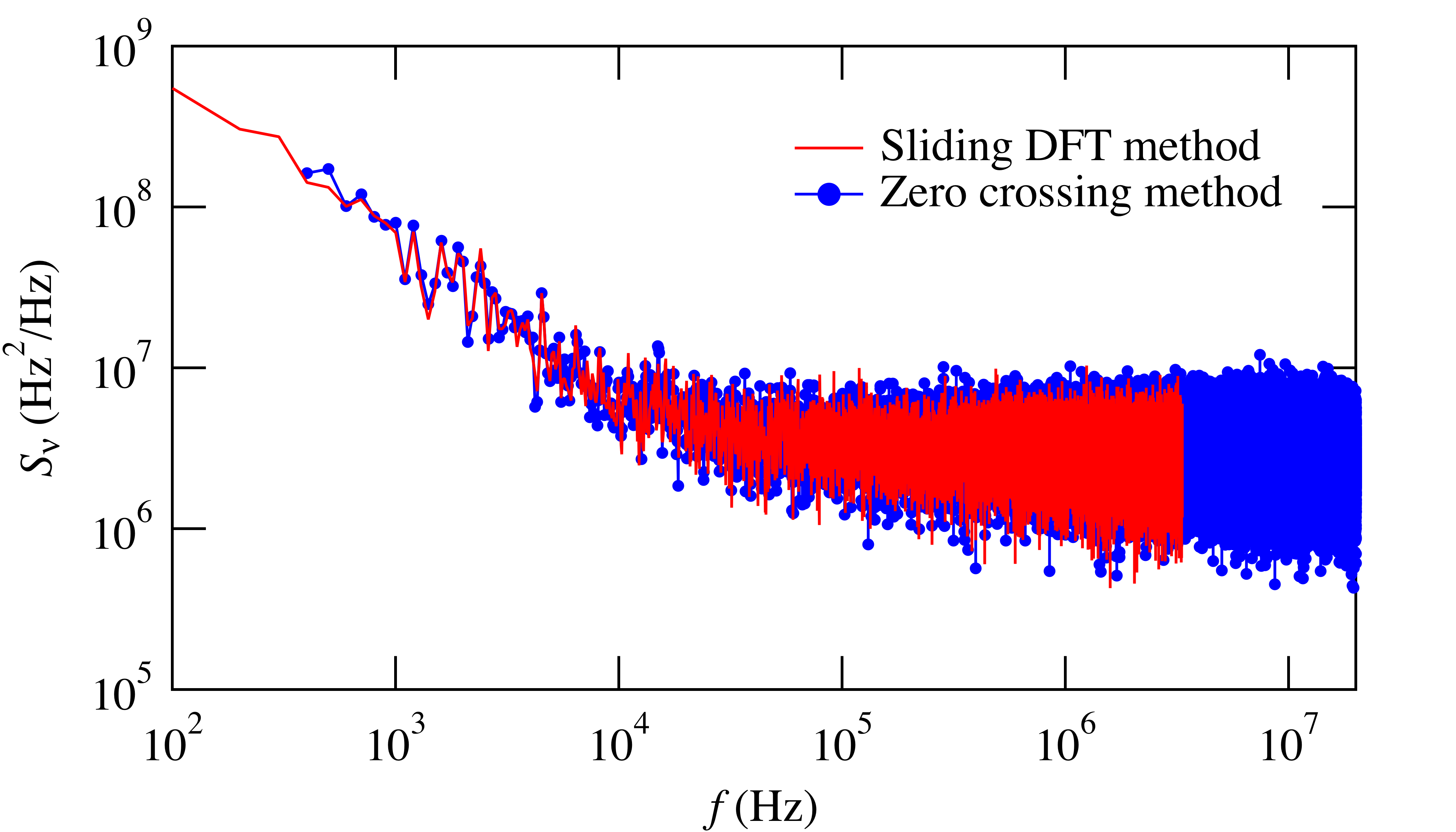}
\par\end{centering}

\caption{(Color online) Comparison of $S_{\mathrm{\nu}}(f)$ from two analysis
methods for the NiFe device. The curves are those for $\unit[10]{ms}$
segments shown in Figures~\ref{Fig:SnuZeros_NiFe}(d) and \ref{Fig:SnuDFT_NiFe}(c),
here shown for all $f\leq\unit[20]{MHz}$. The agreement indicates
that numerical artifacts are likely negligible in both methods.}

\label{Fig:SnuZeros&DFT_NiFe}
\end{figure}

In Fig.~\ref{Fig:SnuZeros&DFT_NiFe} we compare the average $S_{\mathrm{\nu}}(f)$
for $\unit[10]{ms}$ segments obtained from the two analysis methods.
The two methods give nearly identical results over their common frequency
range. Such agreement between two different routes to the same quantity
is evidence that neither method is distorted by numerical artifacts,
and thus that both methods reveal the actual frequency fluctuations
of the oscillator %
\footnote{Comparing the two analysis methods for a variety of experimental conditions
and analysis parameters can reveal each method's limitations. As an
example, for IF waveforms with the smallest signal-to-noise ratios
(smaller than about half that shown in Fig.~\ref{Fig:Vif_vs_Time_NiFe}),
we found that the zero crossing method gave distorted results while
the DFT method remained robust.%
}. Comparing frequency ranges in Fig.~\ref{Fig:SnuZeros&DFT_NiFe},
the zero crossing method extends up to $f=\unit[20]{MHz}$, while
the DFT method is limited to $f\lesssim\unit[3]{MHz}$ by the minimum
DFT segment length mentioned above. On the other hand, since the DFT
method avoids the spurious effects related to segment length, it extends
to lower frequencies than the zero crossing method (for a given segment
length).

\section{Results and Discussion}

\begin{figure}
\begin{centering}
\includegraphics[width=8.5cm]{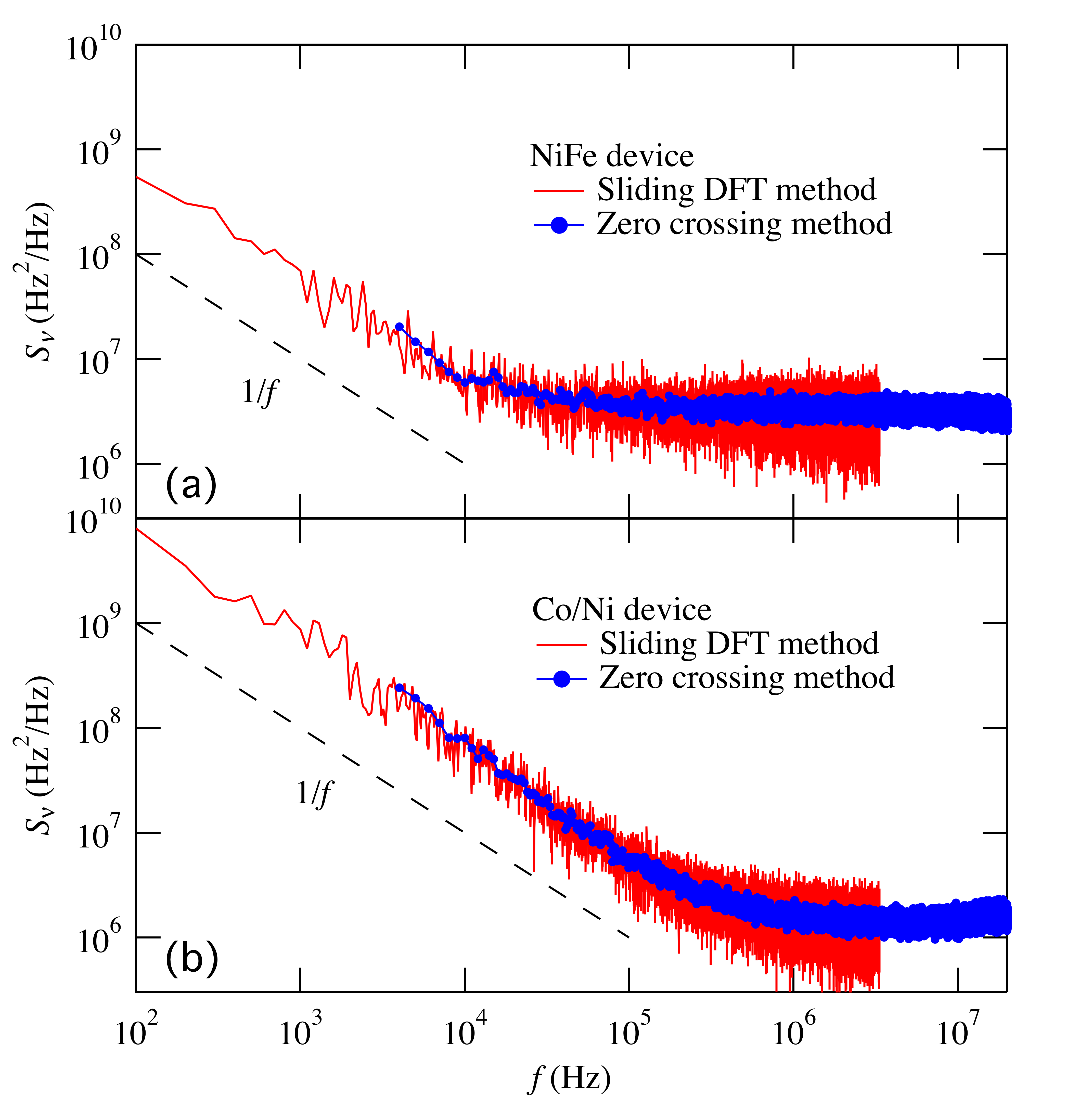}
\par\end{centering}

\caption{(Color online) Combined frequency noise from both analysis methods
for (a) NiFe device and (b) Co/Ni device. Segment length is $\unit[1]{ms}$
for the zero crossing method and $\unit[10]{ms}$ for the sliding
DFT method. Dashed lines are visual guides indicating a $1/f$ power
law spectrum.}

\label{Fig:SnuComposite}
\end{figure}

As mentioned above, we combine spectra from the zero crossing and
DFT methods to obtain $S_{\mathrm{\nu}}(f)$ over a larger frequency
range than is possible with either method alone. As shown in Fig.~\ref{Fig:SnuComposite},
the zero crossing method using $\unit[1]{ms}$ segments together with
the DFT method using $\unit[10]{ms}$ segments yields $S_{\mathrm{\nu}}(f)$
spanning more than five decades in $f$. Both the NiFe and Co/Ni devices
show the same overall behavior: $S_{\mathrm{\nu}}(f)$ is constant
at large $f$ (excluding the rolloff and noise floor features mentioned
above) and varies as approximately $1/f$ at small $f$. We found
this same qualitative behavior in all STOs we measured. For a given
device, we have not found a clear dependence of either the white or
$1/f$ noise on bias current or applied field, but our measurements
to date have consisted of a broad survey rather than a search for
systematic trends. One clear pattern that does emerge from our data
is that the Co/Ni devices have stronger $1/f$ noise than the NiFe
devices, as illustrated by the representative spectra in Fig.~\ref{Fig:SnuComposite}.
The white noise for the two device types is typically comparable (factor
of 2.5 difference in Fig.~\ref{Fig:SnuComposite}), while the $1/f$
component is typically 10 times larger in the Co/Ni devices (factor
of 15 difference in Fig.~\ref{Fig:SnuComposite}). Whether this systematic
difference is due to the different materials in the STO free layer,
the different magnetic anisotropies (which lead to different precessional
trajectories), or to other factors is an important topic for future
investigations.

Several considerations rule out sources for the $1/f$ noise other
than the STO itself. Noise from the bias current source is filtered
by the dc path of the bias tee, which has a bandwidth of $\unit[8]{kHz}$,
whereas we observed $1/f$ noise up to much higher frequencies. Furthermore,
an effect due to bias current noise would scale with $|d\nu_{0}/dI_{\textrm{b}}|$
of the STO, but the devices shown in Fig.~\ref{Fig:SnuComposite}
follow the opposite trend: the NiFe device has a \emph{larger} $d\nu_{0}/dI_{\textrm{b}}$
(by a factor of $\approx3$) but \emph{weaker} $1/f$ noise than the
Co/Ni device. Noise from other sources such as the oscilloscope, preamplifier,
or stray magnetic fields would affect all our measurements equally,
which is not consistent with the reproducible differences among devices
that we observed.

Another trend emerges when we compare the measured SA linewidth, $\Delta\nu_{\textrm{SA}}$,
with the value expected from the white noise level in $S_{\mathrm{\nu}}(f)$,
$\Delta\nu_{\textrm{wh}}$. (At this point we report Lorentzian linewidths
for a measurement time of $\unit[1]{s}$; see below for why it is
important to specify both line shape and time scale.) Our NiFe devices
have SA linewidths that are 1.1 to 1.3 times larger than expected
from the white noise: $\Delta\nu_{\textrm{SA}}=\unit[11.2]{MHz}$
and $\Delta\nu_{\textrm{wh}}=\unit[10.4]{MHz}$ for the device in
Fig.~\ref{Fig:SnuComposite} (all values of $\Delta\nu$ here have
an uncertainty of approximately $\pm\unit[0.5]{MHz}$ unless error
bars on a plot indicate otherwise). In constrast, our Co/Ni devices
have SA linewidths that are 2 to 3 times larger than expected from
the white noise: $\Delta\nu_{\textrm{SA}}=\unit[10.0]{MHz}$ and $\Delta\nu_{\textrm{wh}}=\unit[4.1]{MHz}$
for the device in Fig.~\ref{Fig:SnuComposite}. As we describe next,
this trend can be understood as a consequence of the different $1/f$
noise strength in the different devices.

Spectral line broadening due to $1/f$ frequency noise is well known
in the field of single-mode semiconductor lasers. We first give a
brief description of the key concepts and then apply them to our measurements
in the next paragraph. White frequency noise in a laser, caused by
spontaneous emission events, gives a spectral line having a Lorentzian
shape \cite{Henry:1982fk}. Frequency noise having a $1/f$ spectrum,
caused by the fluctuating number of charge carriers in the semiconductor,
gives an additional contribution to the spectral line having a Gaussian
shape \cite{Mercer:1991zr,Stephan:2005ly}. When the white and $1/f$
contributions to the spectral line are comparable, the shape can be
described by a convolution of Lorentzian and Gaussian profiles known
as a Voigt function \cite{Mercer:1991zr,Stephan:2005ly}. Importantly,
the broadening depends not only on the white and $1/f$ noise strengths,
but also on the \emph{time scale of the measurement}. Semiconductor
laser spectra are typically measured by interfering the light with
a delayed copy of itself at a photodiode detector and modulating one
arm of the interferometer (typically at $\unit[100]{MHz}$) to avoid
low frequency photodiode noise \cite{Okoshi:1980bh}. (The delay is
achieved by placing a length of optical fiber in one arm of the interferometer.)
This method essentially translates spectral lines from optical frequencies
to radio frequencies, where they can be measured with a conventional
SA. The delay time $T_{\textrm{del}}$ in such a measurement sets
a lower limit on the frequency noise: only $S_{\mathrm{\nu}}(f)$
above $f\approx1/T_{\textrm{del}}$ contributes to the spectral line
\cite{Mercer:1991zr}. Once $T_{\textrm{del}}$ is long enough that
this lower limit lies in the $1/f$ region of $S_{\mathrm{\nu}}(f)$,
the spectral line will become broader as $T_{\textrm{del}}$ increases.
A numerical study of cases where the white and $1/f$ noise contributions
were comparable showed that the overall (Voigt) linewidth varies approximately
logarithmically with delay time \cite{Mercer:1991zr}.

To apply the semiconductor laser picture to an electronic oscillator
such as an STO, we must consider the appropriate time scale for an
SA measurement. For a single SA sweep, the analog of $T_{\textrm{del}}$
is the sweep time over which the LO moves through a span around $\nu_{0}$.
The minimum sweep time for commercial instruments is typically $\unit[1]{ms}$.
Comparing this time with the spectra in Fig.~\ref{Fig:SnuComposite},
we see that individual SA sweeps involve times over which the frequency
noise in our STOs is not white. Moreover, several sweeps are usually
averaged together and there is a dead time of $\approx\unit[100]{ms}$
between sweeps (required for restabilization of the LO). We typically
averaged 10 sweeps to produce the final $S_{V}(\nu)$ that we fit
to determine $\Delta\nu_{\textrm{SA}}$, thus the total measurement
time was $T_{\textrm{meas}}\approx\unit[1]{s}$. Certainly frequency
noise below $f\approx\unit[1]{Hz}$ cannot contribute to the spectral
line, but the dead time prevents an exact mapping between $T_{\textrm{del}}$
in the laser measurement and $T_{\textrm{meas}}$ for our averaged
SA measurements. We therefore proceed by developing a numerical model
of a swept SA that simulates the signal processing occuring after
the mixer, \emph{i.e.}, the steps that convert $V_{\textrm{IF}}(t)$
into a spectrum $S_{V}(\nu)$ averaged over multiple sweeps.

\begin{figure}[h]
\begin{centering}
\includegraphics[width=6cm]{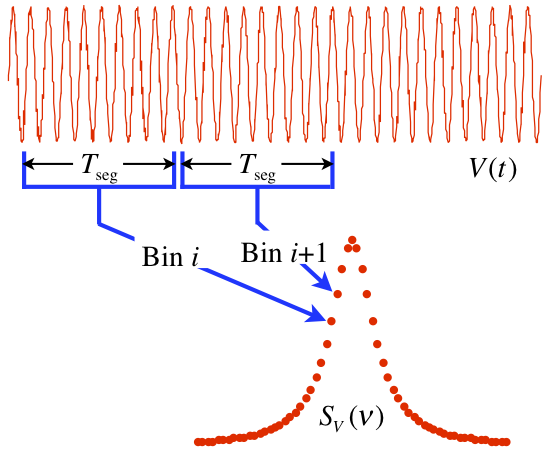}
\par\end{centering}

\caption{(Color online) Illustration of how a swept SA generates a spectrum
from a voltage waveform.}

\label{Fig:SpectAnalyzer}
\end{figure}

The process by which an SA converts a time domain signal into a frequency
domain spectrum, illustrated in Fig.~\ref{Fig:SpectAnalyzer}, includes
various time scales that must be incorporated into a model SA. For
the \emph{i}th frequency bin in the spectrum, the LO frequency of
the mixer is fixed at $\nu_{\textrm{LO}}^{i}$ during a segment time
$T_{\textrm{seg}}$ (although SAs often sweep the LO continuously,
here we consider it to be stepped discretely). The mixer IF output
goes through a bandpass filter, whose width is set by the resolution
bandwidth (RBW) of the SA, and then through a power detector %
\footnote{In an actual SA, the LO is offset from the center of the frequency
bin so that the IF and the bandpass filter are centered around a frequency
that is optimal for subsequent signal processing. This is how our
SA generates the IF signal centered around $\unit[70]{MHz}$ that
we use for our time domain measurements.%
}. The power detected during the \emph{i}th segment is the value of
$S_{V}$ for the \emph{i}th frequency bin. For a spectrum spanning
$n$ frequency bins, the time to acquire a single spectrum is $T_{\textrm{sweep}}=nT_{\textrm{seg}}$.
When $N$ spectra are averaged together, the time to restabilize $\nu_{\textrm{LO}}$
at the first bin is typically much larger than $T_{\textrm{sweep}}$
and thus the total measurement time $T_{\textrm{meas}}$ is much larger
than $NT_{\textrm{sweep}}$. Although this picture omits many details
of the inner workings of an actual SA, it contains the relevant time
scales that determine how non-white frequency noise affects the spectrum.

We created a model SA that takes $V_{\textrm{IF}}(t)$ as the input,
rather than $V(t)$, since this is the data we recorded for our STOs.
Thus rather than changing $\nu_{\textrm{LO}}$ to generate different
frequency bins, we changed the center frequency of the bandpass filter.
After filtering each segment (using a filter with 3~dB points at
half the RBW away from the center frequency and a rolloff of $\unit[36]{dB/decade}$),
we computed the ``power'' in each bin by simply summing the squares
of the values in the segment %
\footnote{Actual SAs are carefully designed to yield accurate power values for
a variety of detection modes. Since we are interested only in the
normalized width and shape of $S_{\mathrm{V}}(\nu)$, we can ignore
many effects that affect the power in all bins equally.%
}. From each set of $n$ segments we obtained a single spectrum, and
we repeated the process to obtain $N$ individual spectra for each
value of $T_{\textrm{seg}}$. We fit the individual $S_{V}(\nu)$
to determine $\Delta\nu_{\textrm{SA}}$ for $T_{\textrm{meas}}=nT_{\textrm{seg}}$
and we fit the average $S_{V}(\nu)$ to determine $\Delta\nu_{\textrm{SA}}$
for $T_{\textrm{meas}}=NnT_{\textrm{seg}}$ (unlike in a real SA,
there is no dead time between sweeps in our model SA). We used primarily
the Voigt function to fit $S_{V}(\nu)$ because it can fit lines that
are Lorentzian, Gaussian, or any mixture of the two. We first present
the Voigt results and then discuss fits using pure Lorentzian and
Gaussian functions.

\begin{figure}[h]
\begin{centering}
\includegraphics[width=8.5cm]{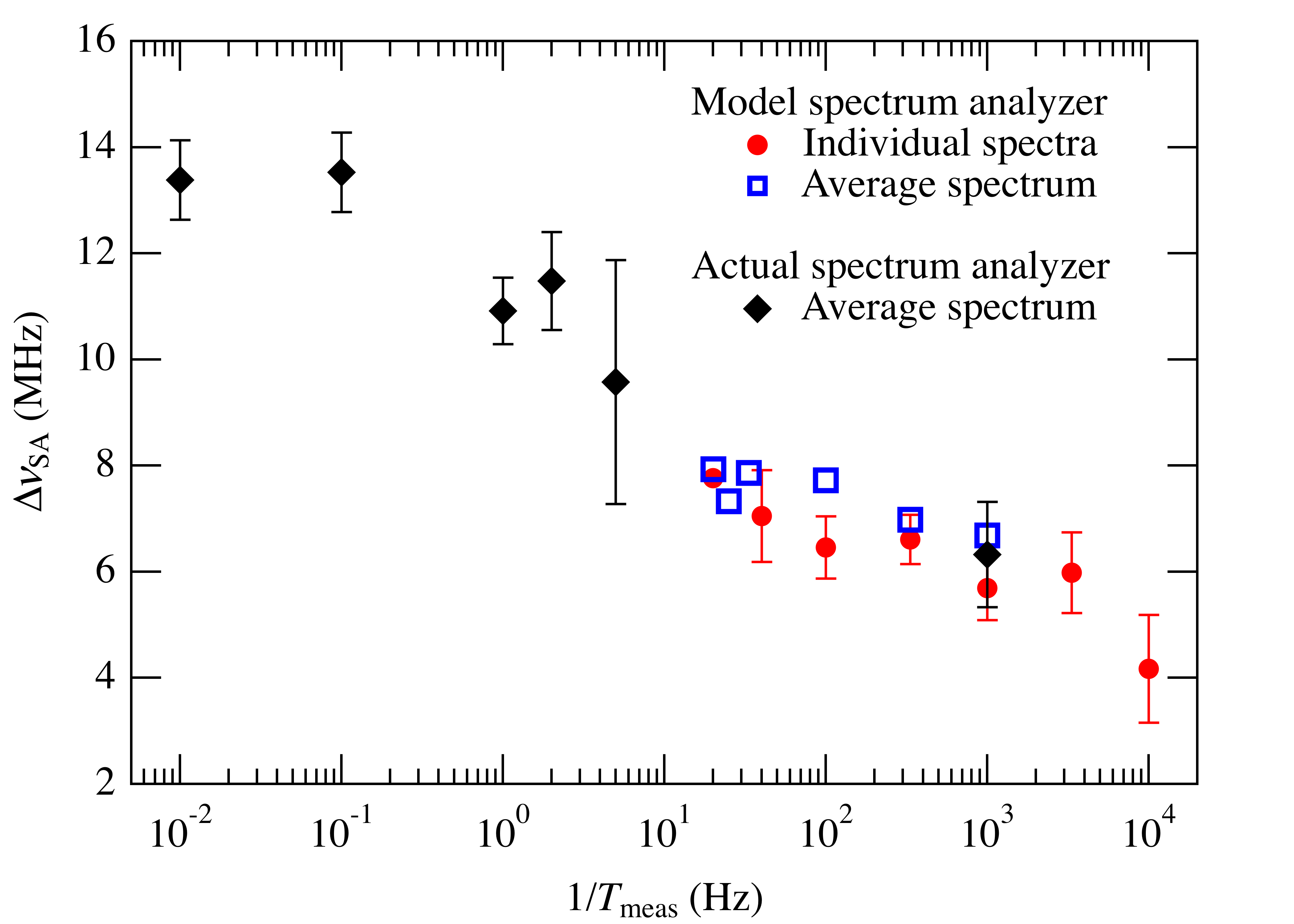}
\par\end{centering}

\caption{(Color online) Dependence of Voigt linewidth on inverse measurement
time for the Co/Ni device. Model SA parameters are $\textrm{RBW}=\unit[1]{MHz}$,
$n=100$, $N=10$ (except for the longest segments where the $\unit[50]{ms}$
IF waveform was used up before reaching 10 spectra). Error bars for
solid circles show the standard deviation of the mean for a set of
$N$ values with the same $T_{\textrm{meas}}$. Actual SA parameters
are $\textrm{RBW}=\unit[1]{MHz}$, $T_{\textrm{sweep}}=\unit[1]{ms}$,
dead time between sweeps $\approx\unit[100]{ms}$. Error bars for
solid diamonds show the standard deviation of the mean for a set of
5 successive measurements with the same $T_{\textrm{meas}}$.}

\label{Fig: FWHM_vs_Tmeas}
\end{figure}

Figure~\ref{Fig: FWHM_vs_Tmeas} shows Voigt linewidth \emph{vs.}
inverse measurement time for the Co/Ni device measured by (1) an actual
SA, and (2) our model SA applied to the same $\unit[50]{ms}$ $V_{\textrm{IF}}(t)$
waveform used to compute $S_{\mathrm{\nu}}(f)$. For the model SA,
we used $n=100$ and varied $T_{\textrm{meas}}$ by choosing values
of $T_{\textrm{seg}}$ between $\unit[1]{\mu s}$ (shorter segments
gave unreliable results) and $\unit[500]{\mu}s$. We report both the
mean from fits to 10 individual spectra and the fit to the average
of these 10 spectra. For the actual SA, we varied $T_{\textrm{meas}}$
by averaging with $N=1$, 2, 5, 10, 100, and 1000, repeating the measurement
five times for each value of $T_{\textrm{meas}}$ in order to report
a mean value and estimate its uncertainty. The actual and model SA
results both show that $\Delta\nu_{\textrm{SA}}$ increases logarithmically
with measurement time, and they agree quantitatively for $T_{\textrm{meas}}=\unit[1]{ms}$.
When applied to the measured $V_{\textrm{IF}}(t)$ for the NiFe device,
the model SA yields a weaker dependence of $\Delta\nu_{\textrm{SA}}$
on measurement time (the change in $\Delta\nu_{\textrm{SA}}$ barely
exceeds the statistical uncertainty over the accessible range of $T_{\textrm{meas}}$),
which is consistent with the weaker $1/f$ noise in $S_{\mathrm{\nu}}(f)$
for this device. We also applied the model SA to numerically generated
signals. For signals having white frequency noise, $\Delta\nu_{\textrm{SA}}$
was independent of measurement time, whereas for signals having $S_{\mathrm{\nu}}(f)$
comparable to that measured for our Co/Ni device, $\Delta\nu_{\textrm{SA}}$
\emph{vs.} $1/T_{\textrm{meas}}$ had a slope similar to that seen
in Fig.~\ref{Fig: FWHM_vs_Tmeas}. Thus the model SA applied to signals
having a range of $S_{\mathrm{\nu}}(f)$ from strictly white to strongly
$1/f$ indicates that the dependence of linewidth on measurement time
is a direct consequence of non-white frequency noise.

\begin{figure}
\begin{centering}
\includegraphics[width=8.5cm]{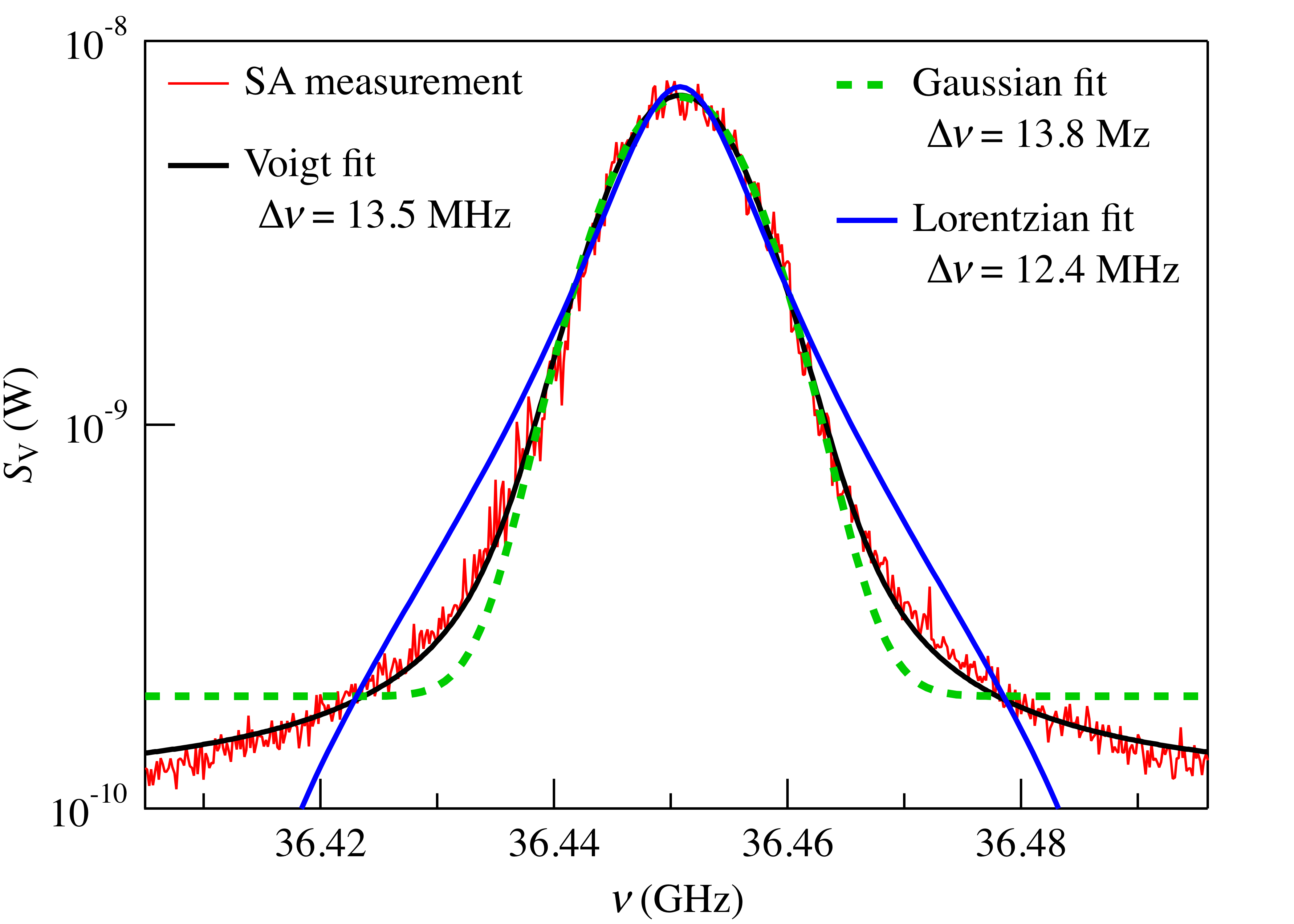}
\par\end{centering}

\caption{(Color online) Spectral line for Co/Ni device with $T_{\textrm{meas}}=\unit[10]{s}$.
Lines show fits to Voigt, Gaussian, and Lorentzian functions.}

\label{Fig: SA_Navg100_Fits}
\end{figure}

As mentioned above, broadening due to $1/f$ frequency noise is accompanied
by a change in the shape of the spectral line \cite{Mercer:1991zr,Stephan:2005ly},
with the Voigt shape changing from mostly Lorentzian to mostly Gaussian
as $T_{\textrm{meas}}$ increases. The noise in our SA data for most
values of $T_{\textrm{meas}}$ is too large to discern this trend
clearly, \emph{i.e.}, a pure Lorentzian shape fits about as well as
a Voigt shape. However, for the CoNi device at $T_{\textrm{meas}}\geq\unit[10]{s}$
a Voigt shape clearly fits the data better than a Lorentzian shape,
as shown in Fig.~\ref{Fig: SA_Navg100_Fits}. We also show a pure
Gaussian fit to this line for comparison, and we see that the Voigt
function provides the best fit to the entire line. The Lorentzian
shape gives a good fit to all spectra from the NiFe device, although
we did not measure beyond $T_{\textrm{meas}}=\unit[1]{s}$ for the
particular oscillation mode presented here. This is consistent with
previous work on similar NiFe devices that did not find significant
deviations from a Lorentzian line shape \cite{Rippard:2004ad,Rippard:2004fr}.

In terms of the overall picture of STO spectral lines, our results
mean there are \emph{two sources for a Gaussian contribution to line
shape}. The first involves the rate of relaxation to the stable precessional
orbit, combined with the dependence of frequency on precessional amplitude
(the {}``frequency nonlinearity'' intrinsic to STOs \cite{Tiberkevich:2008ys}),
which sets a correlation time for phase fluctuations. When this correlation
time is short compared to the thermal dephasing time the line shape
is Lorentzian; in the opposite limit it is Gaussian \cite{Tiberkevich:2008ys}.
In terms of frequency noise, this correlation suppresses $S_{\mathrm{\nu}}(f)$
at large $f$ according to \begin{equation}
S_{\nu}(f)=\frac{S_{\textrm{wh}}}{1+(\nicefrac{2\pi f}{\eta})^{2}},\label{eq:Snu_rolloff}\end{equation}
where $\eta$ is the relaxation rate \cite{SilvaKeller_preprint2010}.
The second source for a Gaussian line shape is $1/f$ frequency noise,
as described above, which can be understood as a correlation at long
times that affects $S_{\mathrm{\nu}}(f)$ at small $f$.

Distinguishing the two mechanisms for a non-Lorentzian STO line shape
clearly requires more than a single SA measurement. Since the relaxation
mechanism does not depend on measurement time, SA measurements over
a wide range of $T_{\textrm{meas}}$ could indicate whether the $1/f$
mechanism is significant, but a quantitative conclusion about the
relative contributions of the two mechanisms would require considerable
care. Another approach is to use the autocorrelation function of the
STO signal to measure the correlation at short times, as done in recent
experimental work \cite{Boone:2009fk} where the deviation from a
Lorentzian line shape was less pronounced than that seen in Fig.~\ref{Fig: SA_Navg100_Fits}.
Finally, both the relaxation rate and $1/f$ noise can be seen directly
in $S_{\mathrm{\nu}}(f)$ if it is measured over a large enough range
of $f$. This approach has the advantage that each mechanism can be
quantified separately.

As mentioned above, in our measurements the limited IF bandwidth suppressed
$S_{\mathrm{\nu}}(f)$ above $f\approx\unit[20]{MHz}$. While this
prevents a direct measurement of $\eta$, it does set a lower bound
of $\eta/2\pi\approx\unit[20]{MHz}$ (see Eq.~\eqref{eq:Snu_rolloff}).
From the theory of the relaxation mechanism \cite{Tiberkevich:2008ys,SilvaKeller_preprint2010},
the condition for a negligible Gaussian contribution can be written
as $\eta/2\pi\gg\Delta\nu_{\textrm{wh}}$. Since the measured values
of $\Delta\nu_{\textrm{wh}}$ ($\unit[10.4]{MHz}$ for the NiFe device;
$\unit[4.1]{MHz}$ for the Co/Ni device) are less than $\unit[20]{MHz}$,
we can conclude that the line shape for both devices would be close
to Lorentzian in the absence of $1/f$ noise. The strongly non-Lorentzian
line in Fig.~\ref{Fig: SA_Navg100_Fits} for the Co/Ni device can
be unambiguously attributed to $1/f$ frequency noise.

Although non-white frequency noise has not been directly measured
in previous STO experiments, recent reports indicate it is probably
an important effect in devices beyond the two types considered here.
The line ``jitter'' and SA linewidth increasing with measurement time
reported in MgO nanopillar STOs \cite{Devolder:2009ve} can both be
interpreted as evidence for $1/f$ frequency noise, and the authors
suggest a possible mechanism for such noise that is specific to their
particular devices. In other nanopillar STOs containing MgO \cite{Houssameddine:2009zm}
or metallic \cite{Pribiag:2009by} barriers, the DFT of selected short
segments yielded linewidths much smaller than those found from either
DFT or SA measurements averaged over long times. Beyond these published
reports, we and others have noticed when watching the non-averaged
SA display that some devices show larger trace-to-trace jumps in center
frequency than others, an observation that may be explained by varying
amounts of $1/f$ noise in the devices.

\section{Conclusion}

We measured the output of two types of STOs in the time domain, using
the IF ouput of an SA to access signals well above 10~GHz. We presented
two techniques for obtaining the power spectrum of frequency noise,
$S_{\mathrm{\nu}}(f)$, and showed the advantage of combining them
to yield an averaged $S_{\mathrm{\nu}}(f)$ over a wide range of $f$.
The $1/f$ noise we observed indicates that theoretical models based
on thermal noise sources are insufficient for our devices over times
longer than about $\unit[1]{\mu s}$. We also measured spectral linewidths
in our devices using both an actual SA and a numerical model that
allowed shorter measurement times. For the devices with the strongest
$1/f$ noise, we found that SA linewidth increased, and the line shape
became significantly non-Lorentzian, as measurement time increased.
These results imply that SA measurements of STOs should be accompanied
by measurement time values so that (1) comparisons can be made among
various STOs measured by different researchers, and (2) non-Lorentzian
line shapes can be correctly interpreted. Although the consequences
of our results for STO applications depend on the time scales involved,
we expect that measurements of $S_{\mathrm{\nu}}(f)$ will allow more
accurate predictions of performance than SA measurements alone for
many applications. The detailed picture of oscillator noise provided
by $S_{\mathrm{\nu}}(f)$ will also help in distinguishing among different
physical origins for the noise.

\end{document}